\documentclass[%
 reprint,
 amsmath,amssymb,
 aps,
]{revtex4-2}

\pdfoutput=1 
\usepackage{graphicx}
\usepackage{dcolumn}
\usepackage{bm}

\usepackage{color}
    \definecolor{darkgreen}{rgb}{0,0.5,0}
    \definecolor{darkred}{rgb}{0.5,0,0}
    \definecolor{darkblue}{rgb}{0,0,0.6}
    \definecolor{purple}{rgb}{0.4,.2,0.7}

\begin{document}

\preprint{APS/123-QED}

\title{Rotating black holes in Randall-Sundrum II braneworlds}

\author{William~D.~Biggs}
\author{Jorge~E.~Santos}%
\affiliation{%
 DAMTP, University of Cambridge
}%

\date{\today}

\begin{abstract}
We find rotating black hole solutions in the Randall-Sundrum II (RSII) model, by numerically solving a three-dimensional PDE problem using pseudospectral collocation methods. We compute the area and equatorial inner-most stable orbits of these solutions. For large black holes compared with the AdS length scale, $\ell$, the black hole exhibits four-dimensional behaviour, approaching the Kerr metric on the brane, whilst for small black holes, the solution tends instead towards a five-dimensional Myers-Perry black hole with a single non-zero rotation parameter aligned with the brane. This departure from exact four-dimensional gravity may lead to different phenomenological predictions for rotating black holes in the RSII model to those in standard four-dimensional general relativity. This letter provides a stepping stone for studying such modifications.
\end{abstract}

\maketitle
\section{Introduction}
For some time there has been great interest in the idea that our Universe is a brane embedded in a higher dimensional space. In most such theories, the extra dimensions are compact and very small, so that four-dimensional physics is reproduced. The five-dimensional RSII braneworld model \cite{Randall:1999vf}, on the other hand, contains an extra dimension that is non-compact but warped so that the low-energy behaviour of gravity on the brane still yields four-dimensional general relativity. The RSII model and its predecessor, RSI \cite{Randall:1999ee}, were first introduced as a mechanism for solving the hierarchy problem: due to the warping factor, the effective mass of an object in the four-dimensional theory is exponentially suppressed compared with its proper five-dimensional mass, leading to a large hierarchy between gravity and the other forces. Whilst matter fields may be restricted to the brane, gravity (at least at higher energies) must be able to propagate through all dimensions. This motivates finding black hole solutions in the RSII model, since these may yield different phenomenological predictions to ordinary four-dimensional general relativity.

Initially there was some debate over whether stable, large black holes on the brane could exist. Due to arguments from the AdS/CFT correspondence \cite{Maldacena:1997re,Witten:1998qj,Aharony:1999ti}, the low-energy theory on the brane can be described by gravity coupled to a large $N$, strongly coupled CFT \cite{Verlinde:1999fy,Gubser:1999vj, Hawking:2000kj}. Therefore a black hole in the five-dimensional RSII bulk would correspond holographically to a \textit{quantum-corrected} black hole on the brane \cite{Emparan:2002px}. It was thought that the black hole would quickly evaporate due to extra radiation arising from the CFT degrees of freedom, meaning any bulk black hole solution would be time-dependent. However, it was contended that this argument neglects the strong coupling of the CFT \cite{Fitzpatrick:2006cd}. The debate was settled in \cite{Figueras:2011gd}, where static, stable black hole solutions on the brane were found for both small and large radii. This solution is closely related to black droplets and funnels \cite{Hubeny:2009ru,Hubeny:2009kz,Hubeny:2009rc,Caldarelli:2011wa,Figueras:2011va,Santos:2012he,Fischetti:2012ps,Fischetti:2012vt,Santos:2014yja,Fischetti:2016oyo,Marolf:2019wkz,Santos:2020kmq}, which are the gravitational duals to the limit where the CFT decouples from gravity.

In order for the RSII model to be phenomenologically viable it must admit not only static black hole solutions, but also rotating ones. In this paper we present the first fully backreacted, rotating black hole solution in the RSII model, by generalising the method pioneered in \cite{Headrick:2009pv}. We were able to find solutions over a two parameter space. One parameter runs over the possible angular velocity of the black hole, and we were able to find solutions close to extremality. The second parameter controls the size of the black hole relative to the five-dimensional AdS length, $\ell$. We found that for large rotating RSII black holes, the induced metric on the brane closely resembles the four-dimensional Kerr black hole, whilst small rotating RSII black holes exhibit five-dimensional behaviour, approaching the five-dimensional Myers-Perry black holes \cite{Myers:1986un} with a single non-zero rotation aligned with the brane.

This transition from four-dimensional to five-dimensional behaviour means that finite-sized RSII black holes have slightly different induced geometry on the brane to the usual four-dimensional Kerr black hole, and thus will lead to different phenomenological predictions for four-dimensional observers. Since the black holes in our Universe are expected to be rotating and uncharged, this provides the possibility of testing the RSII model via astrophysical measurements of black holes as well as in future lepton colliders, such as the International Linear Collider \cite{ILC} and the Compact Linear Collider \cite{Roloff:2019crr}. For example, we show that the radius of the inner-most stable equatorial circular orbits (ISCOs) predicted by our solutions differ slightly to that of the Kerr black hole of the same rotation.
\section{Constructing rotating RSII \\ black holes}\label{Construct}
We seek a solution to the five-dimensional Einstein equation with negative cosmological constant,
\begin{equation}
R_{ab}+\frac{4}{\ell^2}g_{ab}=0\,,
\label{eq:einstein}
\end{equation}
where $\ell$ is the five-dimensional AdS radius. To motivate our choice of coordinates, we first consider AdS$_5$ in Poincar\'e coordinates:
\begin{equation}
\label{AdS5}
    g_{AdS_5} = \frac{\ell^2}{z^2}\left[-\mathrm{d}t^2+\mathrm{d}z^2+\mathrm{d}r^2+r^2 \left(\mathrm{d}\theta^2+\sin^2\theta\,\mathrm{d}\varphi^2\right)\right]\,,
\end{equation}%
where the last two terms correspond to the line element of the round two-sphere with $\theta\in(0,\pi)$, $\varphi\sim \varphi+2\pi$ being the standard polar and azimuthal angles, respectively.

The RSII model with no matter can be thought of as a part of the Poincar\'e patch of AdS between the Poincar\'e horizon and a constant $z$ slice (the brane), with a $\mathbb{Z}_2$ reflection symmetry across the brane. This motivates taking the transformations,
\begin{eqnarray}
    z = \frac{\Delta(x,y)}{1-y^2}, \quad r = \frac{x\sqrt{2-x^2}}{1-y^2},
\end{eqnarray}
where $\Delta(x,y) = 1-x^2 + \beta^{-1}(1-y^2)$. Now the coordinates $x \in (0,1), y \in (0,1)$ parametrise the space between a brane at $x=1$ (which is a surface of constant $z=\beta^{-1}$), and the Poincar\'e horizon at $y=1$, and so are a natural set of coordinates to find solutions on the RSII brane. The metric becomes
\begin{eqnarray}
\label{RSII}
    g_{\mathrm{AdS}_5} = \frac{\ell^2}{\Delta(x,y)^2}\Bigg[&&-h(y)^2dt^2+\frac{4y^2}{h(y)^2}dy^2  \nonumber\\&& +\frac{4}{g(x)}dx^2+x^2 g(x)d\Omega^2_{(2)}\Bigg],
\end{eqnarray}
where, here and throughout, $g(x) = 2-x^2$ and $h(y)=1-y^2$. These were the coordinates used by \cite{Figueras:2011gd} to find a static black hole on the brane. In our case, we seek to find a rotating black hole on the brane, and hence our solution will be neither static nor spherically symmetric. However, we do assume that the solution will be stationary and axisymmetric (with Killing vector fields $\partial_t$ and $\partial_\phi$ respectively), and that it will be invariant under the transformation of simultaneously swapping the signs of $t$ and $\phi$. The most general \emph{Ansatz}, with a bifurcate Killing horizon at $y=0$, satisfying these assumptions is
\begin{widetext}
\begin{eqnarray}\label{ansatz}
ds^2 = \frac{\ell^2}{\Delta(x,y)^2}\Bigg\{ -&&\frac{h(y)^2 y^2 P(y)}{A(y,u)}F_1\mathrm{d}t^2 + \frac{4 F_2}{P(y)h(y)^2}\mathrm{d}y^2 + \frac{4\,F_5}{g(x)}\left(\mathrm{d}x + \frac{F_8}{h(y)}\mathrm{d}y+F_9\,\mathrm{d}u\right)^2 \nonumber\\
&& +\,x^2 g(x)\left[\frac{4F_3}{2-u^2}\left(\mathrm{d}u+\frac{F_7}{xh(y)}\mathrm{d}y\right)^2 + (1-u^2)^2 S(x,y,u)F_4\left(\mathrm{d}\phi+y^2 W(y,u)F_6\,\mathrm{d}t\right)^2\right]\Bigg\},
\end{eqnarray}
\end{widetext}
where the coordinate $u$ measures the inclination from the equatorial plane and, in pure AdS$_5$, is related to the usual polar angle on the $S^2$ via $u(2-u^2)^{1/2} = \cos{\theta}$. The unknown functions $F_i$ for $i=1,...\,,9$ depend on $\{x\,,y\,,u\}$, whilst $A,\,P,\,S$ and $W$ are known functions, given in the appendix, depending on a parameter $\alpha$, which controls the horizon angular velocity of the braneworld black hole. The coordinate domain is the cube $\{x\,,y\,,u\}\in (0,1)^3$.
\paragraph{The DeTurck Method.}
In order to solve the Einstein equation (\ref{eq:einstein}) we use the DeTurck trick \cite{Headrick:2009pv,Wiseman:2011by,Dias:2015nua}. Effectively, we fix the gauge by adding on a term and solve the resulting equation, which is usually called the \textit{Einstein-DeTurck equation} or the harmonic Einstein equation:
\begin{equation}\label{EDT}
    R_{ab}+\frac{4}{\ell^2}g_{ab} - \nabla_{(a}\xi_{b)} = 0\,,
\end{equation}
where the DeTurck vector is defined by $\xi^a = g^{cd}\left[\Gamma^a_{cd}(g)-\Gamma^a_{cd}(\Bar{g})\right]$. Here, $\Gamma^a_{cd}(\mathfrak{g})$ is the Christoffel connection associated to a metric $\mathfrak{g}$, and $\Bar{g}_{ab}$ is a reference metric which we are free to choose. After adding on this gauge fixing term, demanding our symmetry restrictions and imposing appropriate boundary conditions, the Einstein-DeTurck equation becomes a set of elliptic partial differential equations.

Note, however, that in order for a solution to the Einstein-DeTurck equation to also yield a solution to Einstein's equation, we need $\xi^a = 0$ on solutions of (\ref{EDT}). A solution with non-zero $\xi^a$ is called a \textit{Ricci soliton}. It has been shown in certain cases that Ricci solitons cannot exist \cite{Figueras:2011va,Figueras:2016nmo}, however our work does not satisfy the required conditions to make this claim \textit{a priori}. Instead, we simply solve the Einstein-DeTurck equation and check afterwards that the DeTurck vector is small and tending towards zero in the continuum limit. Since the problem we are solving is elliptic, we know that solutions with nonzero $\xi$ cannot be arbitrarily close to a solution of Einstein's equation. As such we use the norm $\xi^2$ to monitor whether we are approaching a true solution of the Einstein equation (\ref{eq:einstein}).

In our case, we choose the reference metric to be defined by the above \emph{Ansatz} with
\begin{eqnarray}\label{reference}
    F_i(x,y,u) = \begin{cases}
1 \quad &\text{for}\; 1\leq i \leq 5\\
\frac{\alpha}{1+\alpha^2}\;&\text{for}\; i=6\\
0 \quad &\text{for}\; 7\leq i \leq 9.
\end{cases}
\end{eqnarray}
\paragraph{Boundary Conditions.}
Firstly, let us consider the fictitious boundaries of our domain. These consist of the equatorial plane at $u=0$, the north pole at $u=1$, the axis of symmetry where the $S^2$ shrinks to zero size at $x=0$ and the bifurcating Killing horizon at $y=0$. The boundary conditions at each of these fictitious boundaries, which can be found in the appendix, are fixed by requiring regularity.

At the $y=1$ boundary, we require that the metric approaches the Poincar\'e horizon of AdS$_5$. In order to achieve this, we enforce that the metric is equal to the reference metric at this boundary. Since $A,\,P,\,S$ and $W$ are all equal to one at $y=1$, note that this boundary condition means the metric matches (\ref{RSII}) as $y\to 1^-$ after the identification, $\mathrm{d}\varphi = \mathrm{d}\phi + \alpha/(1+\alpha^2)\,\mathrm{d}t$.

Finally we need to consider the boundary conditions on the brane (located at $x=1$). In the RSII set-up, the two sides of the brane are identified under a $\mathbb{Z}_2$ symmetry. Since we are interested in the case where the stress energy tensor on the brane vanishes, the corresponding  \textit{Israel junction conditions} \cite{Israel:1966rt} read
\begin{equation}
    0 = K_{ab} - K \gamma_{ab} + \frac{3}{\ell}\gamma_{ab},
\end{equation}
where $\gamma_{ab}$ is the induced metric on the brane and $K_{ab} = \gamma_{ac} \nabla^c n_b$ is the extrinsic curvature, with $n_b$ being the inward unit normal to the brane. These provide six boundary conditions on the brane. We additionally impose that $\xi_x = 0$ and that $F_8 = F_9 = 0$ at $x=1$, making a total of nine boundary conditions, imposed on the nine functions $F_i$.
\paragraph{Numerics.}
We approximate the PDEs by a set of non-linear algebraic equations defined on a Gauss-Lobatto grid using pseudospectral collocation methods. The resultant algebraic equations are then solved using a standard Newton-Raphson algorithm (see for instance \cite{Dias:2015nua} for a comprehensive review of such methods).

Our \emph{Ansatz} depends on two parameters: $\alpha$ and $\beta$. The parameter $\alpha$ controls the rotation of the black hole, with $\alpha=1$ being an extremal black hole and $\alpha=0$ the static braneworld black hole of \cite{Figueras:2011gd}. The size of the black hole relative to the AdS length scale $\ell$ is in turn controlled by $\beta$. More specifically, the temperature and angular velocity of the braneworld black hole, measured with respect to inertial coordinates at spatial infinity of the braneworld, is given by $T=(4\pi\, \ell\, \beta)^{-1}(1-\alpha^2)/(1+\alpha^2)$ and $\Omega = \alpha\, (\ell\,\beta)^{-1}/(1+\alpha^2)$, respectively. Note that the ratio $\Omega/T$ is independent of $\beta$ and $\ell$.
\section{Results and Discussion}\label{Results}

\begin{figure}[h]
\includegraphics[width=\linewidth]{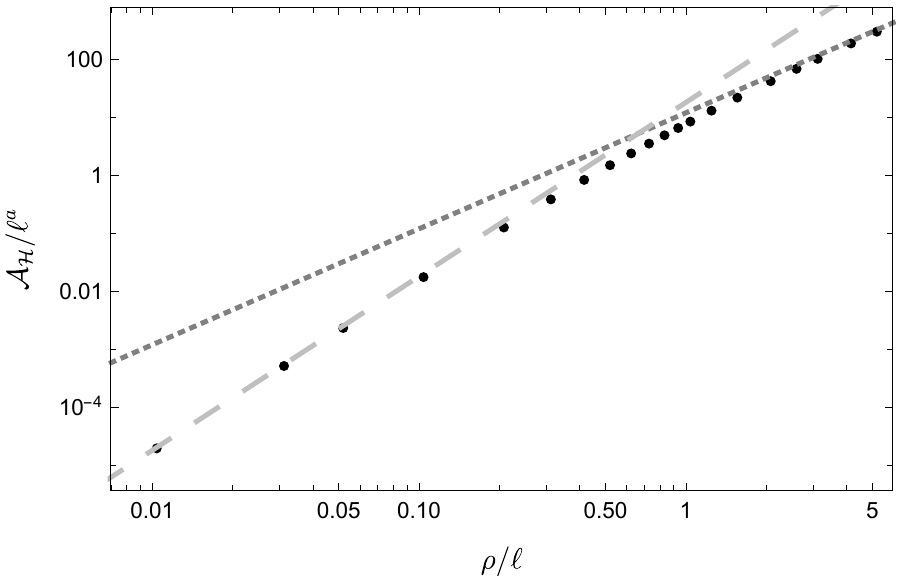}
\caption{\label{fig:entropy} The area of the bifurcating Killing surface of rotating RSII black holes, with $\alpha = 0.2$, as a function of the proper radius $\rho$ (black dots) in a $\log\text{-}\log$ plot. The darker, dotted line is the area of a four-dimensional Kerr black hole, whilst the lighter, dashed line is the area of a five-dimensional single rotating Myers Perry black hole,  both of which have the same $\Omega/T$. We have divided the quantities on both axes by powers of $\ell$ to make them dimensionless, for example on the $y$-axis, we have $\mathcal{A}_\mathcal{H}/\mathcal{\ell}^a$ where $a=3$ for both the RSII black holes and the Myers-Perry black holes, whereas $a=2$ for the Kerr black holes.}
\end{figure}

In Fig.~\ref{fig:entropy} we plot the area of the bifurcating Killing sphere of rotating RSII black holes for a given value of $\alpha$ (or equivalently, $\Omega/T$) against the proper radius, $\rho$ (shown as the black dots in Fig.~\ref{fig:entropy}). To compute $\rho$, we divide the proper distance around the equator of the braneworld black hole by $2\pi$.

For reference, we also plot the areas of asymptotically flat five-dimensional single rotation Myers-Perry black holes (dashed, light grey line) and four-dimensional Kerr black holes (dotted, dark grey line) with the same $\Omega/T$ and appropriately scaled by powers of $\ell$, so that we only compare dimensionless quantities. For small black holes the area exhibits five-dimensional behaviour, whilst large RSII black holes show four-dimensional behaviour.

\begin{figure}[h]
\includegraphics[width=\linewidth]{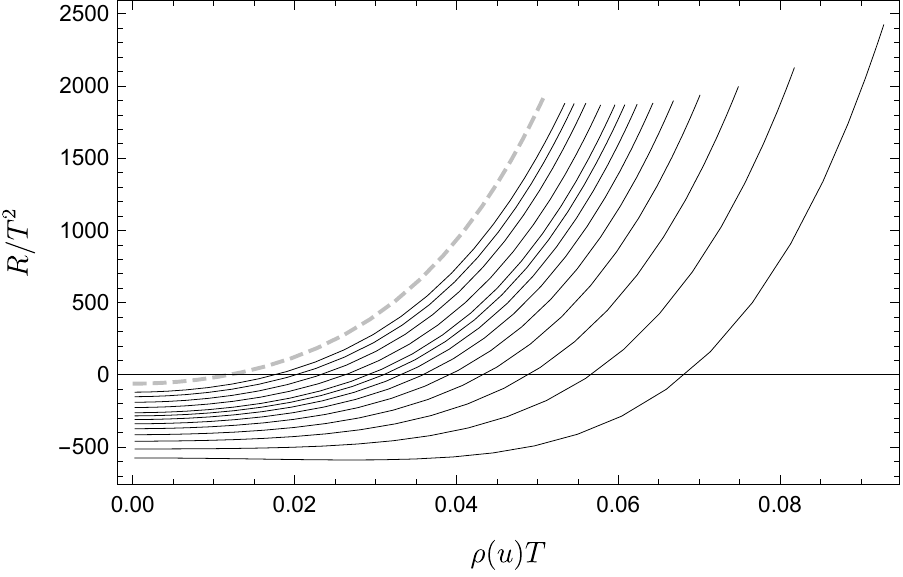}
\caption{\label{fig:ricci} The value of the Ricci scalar of the two-dimensional induced geometry of the intersection of the event horizon with the brane at constant time is plotted here for the braneworld black holes (solid, black lines) and for a four-dimensional Kerr black hole (dashed, grey line) at fixed angular velocity, $\alpha=0.6$. The braneworld curves closest to the Kerr line are those with largest $\beta$, and $\beta$ decreases monotonically as one moves between the curves from left to right.}
\end{figure}

We also consider the induced geometry of the braneworld bifurcating Killing surface, \textit{i.e.} where $x=1$ and $y=0$, at a constant time slice. This yields an axisymmetric two-dimensional geometry, which is completely determined by the Ricci scalar as a function of the inclination from the equatorial plane. In order to express the inclination in a gauge independent way, we introduce the function $\rho(u)$ which measures the proper radius of the circle in the two dimensional geometry at fixed $u$, so that $\rho(1)=0$ at the north pole and $\rho(u)$ increases as $u$ decreases. In Fig.~\ref{fig:ricci} we plot the Ricci scalar as a function of $\rho(u)$ over $u \in (0,1)$ for each of the braneworld black holes (solid, black lines) for a fixed value of $\Omega/T$, along with the corresponding plot for the Kerr black hole (dashed, grey line). Here, and for the remainder of the paper, we multiply the quantities of both axes by the required factors of the temperature, $T$, to make them dimensionless. As one moves from left to right between the different black curves in Fig.~\ref{fig:ricci}, $\alpha$ is fixed and $\beta$ is decreasing.  The curves approach that of the Kerr black hole as $\beta$ becomes large, adding further evidence that the induced geometry of large braneworld black holes is very close to that of the Kerr geometry.

Another interesting quantity that allows for comparison to Kerr is the position of the inner-most stable equatorial circular orbit (ISCO) around the black hole. We consider the motion of a massive particle restricted to the intersection of the equatorial plane $u=0$ and the brane $x=1$. Indeed, both of these planes are at the centre of a $\mathbb{Z}_2$ symmetry in the bulk spacetime, so any geodesic starting with motion in this region will remain within it. Moreover, just like in Kerr, we have two conserved quantities associated to the two Killing vector fields: the energy, $E$ and the angular momentum, $h$. As usual, the use of these allows one to find an ordinary differential equation that governs the radial profile of a normalised timelike geodesic, \emph{i.e.} $y(t)$, which can be written as $\dot{y}(t)^2 = V(y;E,h)$.

To find the ISCO, we require that $V = V' = 0$, where the derivative is with respect to $y$. These two equations can be solved to give a family of circular geodesics depending on $(E,h)$. From these, we pick the one with the minimal angular momentum, which will be the ISCO (minimising with respect to energy gives very similar results). We computed the proper radius $\rho_{ISCO}$ of the ISCO, which once again is the proper distance around its circumference divided by $2\pi$, and multiplied by the temperature $T$ to get a dimensionless quantity. It should be noted that this method requires some interpolation between our lattice points, so will contain some noise.

\begin{figure}[b]
\includegraphics[width=\linewidth]{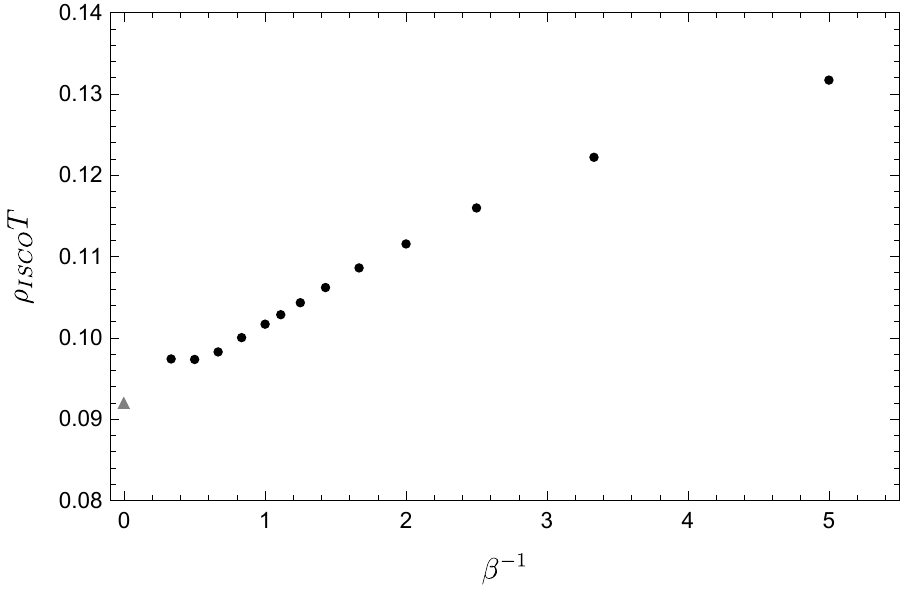}
\caption{\label{fig:isco} The plot of the proper radius of the ISCO for braneworld black holes with $\alpha=0.5$ against $\beta^{-1}$ (black dots). We have added the corresponding value for the Kerr black hole with the same angular velocity at $\beta^{-1}=0$ (grey triangle).}
\end{figure}

In Fig.~\ref{fig:isco}, we plot the value of $\rho_{ISCO}T$ against $\beta^{-1}$ for fixed $\alpha=0.5$. We expect that in the large $\beta$ limit, the braneworld black hole tends towards a Kerr geometry, thus we've added for reference the value for $\rho_{ISCO}T$ for a Kerr black hole of the same angular velocity at $\beta^{-1}=0$. For each value of $\alpha$ we have probed, we observed the same qualitative behaviour: $\rho_{ISCO}T$ tends towards that of the Kerr black hole as $\beta$ becomes large, and increases as $\beta$ decreases. Note that, five-dimensional black holes do not contain bound orbits for timelike particles, and therefore, in agreement with our data, one would expect the value of $\rho_{ISCO}T$ to increase as $\beta$ becomes very small, where our black holes exhibit five-dimensional behaviour.

We also determined the ergoregion of the RSII black holes in the bulk and found in each case that the region has spherical topology, as perhaps expected. It would be interesting to investigate whether the slightly different ergoregions for the black holes result in different superradiance instabilities \cite{Zouros:1979iw,Detweiler:1980uk}.

To conclude, our results clearly show a transition from five-dimensional to four-dimensional behaviour as braneworld black holes increase in size, and hence that rotating braneworld black holes of finite size will differ from the standard four-dimensional Kerr black holes of general relativity. Our results discriminate in a quantitative manner the aforementioned deviations. The work presented in this letter provides a stepping stone for imposing constraints from both multi-messenger astrophysics \cite{Meszaros:2019xej} and future lepton colliders, such as the International Linear Collider \cite{ILC} and the Compact Linear Collider \cite{Roloff:2019crr}, to Randall-Sundrum type scenarios. Natural follow up work includes calculating the quasi-normal mode spectrum of braneworld black holes and studying their linear mode stability. Perhaps more ambitiously, we would like to know whether the braneworld black holes we have found can be formed dynamically on the brane, following the seminal work of \cite{Wang:2016nqi}. Finally, it would be interesting to analyse the extremal limit directly, and compare to the results reported in \cite{Kaus:2009cg}, which studied extremal charged black holes in RSII braneworld scenarios.

\paragraph*{Acknowledgments:}
We would like to thank Harvey~S.~Reall and \'Oscar~C.~Dias for reading an earlier version of this letter and for providing critical comments. J.~E.~S. has been partially supported by STFC consolidated grants ST/P000681/1, ST/T000694/1 and W.~B. was supported by an STFC studentship and a Vice Chancellor's award from the University of Cambridge. The numerical component of this study was carried out using the computational facilities of the Fawcett High Performance Computing system at the Faculty of Mathematics, University of Cambridge, funded by STFC consolidated grants ST/P000681/1, ST/T000694/1 and ST/P000673/1.

\newpage

\appendix

\section*{Supplemental material}

\section{Convergence properties}
\begin{figure}[h!]
\includegraphics[width=\linewidth]{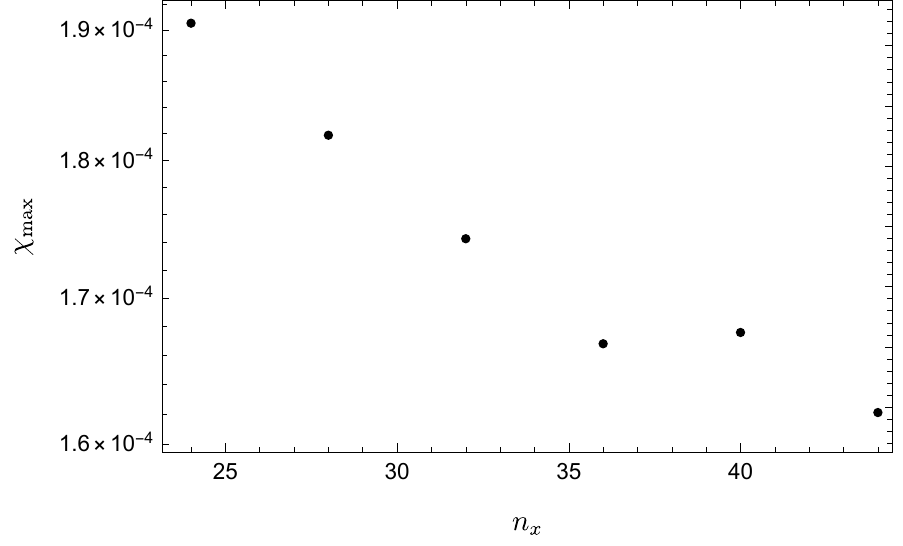}
\caption{\label{fig:conv} The maximum value of the norm of the DeTurck vector, $\chi = \xi^a\xi_a$ against the number grid points in the $x$ direction, denoted by $n_x$. In each case we kept the ratio $n_x:n_y:n_z$ fixed at $4:3:3$.}
\end{figure}

We need to consider the fact that our solutions may be Ricci solitons. We compute the maximum value of the norm of the DeTurck vector, $\chi = \xi^a \xi_a$ across our data points, and seek to show that this value is converging to zero in the continuum limit. In Fig.~\ref{fig:conv}, we have plotted the maximum value of $\chi$ for a given solution ($\alpha=0.9$ and $\beta=1.0$) against the number of lattice points used in the $x$ axis. 

For the majority of our parameter space, the maximum value of $\chi$ is very small, for example, it is of the order of $10^{-4}$ for the case given in Fig.~\ref{fig:conv}. Spectral collocation methods should exhibit exponential convergence as the number of data points is increased, and though we do find that $\chi_{\text{max}}$ decreases with increasing number of points, the convergence is extremely slow. However, this is still consistent with exponential convergence, just one with an extremely small factor multiplying the exponent. The only solutions which have reasonably large values of $\chi_{\text{max}}$, for our resolution of $40\times30\times30,$ are the small black holes with $\beta$ very small. However, these are far less phenomenologically interesting than the large braneworld black holes that more closely resemble four-dimensional black holes. Similar convergence plots have been obtained in \cite{Figueras:2011gd} and \cite{Stein:2016bnr}.

\section{Some explicit expressions}
Recall that the metric (\ref{ansatz}) depends upon four known functions, $A(y,u)$, $P(y)$, $S(x,y,u)$ and $W(y,u)$, each of which depends on the rotation parameter $\alpha$. The explicit forms of these functions are as follows:
\begin{widetext}

\begin{eqnarray}
    A(y,u)&&=1+(1-y^2)^2\alpha^2(2+\alpha^2-y^2(1+\alpha^2-u^2(2-u^2)(1-(1-y^2)\alpha^2)))  \\ \nonumber\\
    P(y)&&=1-(1-y^2)\alpha^2 \\ \nonumber\\
    S(x,y,u)&&= \frac{1+(1-y^2)^2\alpha^2x^2(2+\alpha^2x^2-y^2(1+\alpha^2x^2-u^2(2-u^2)(1-(1-y^2)\alpha^2x^2)))}{(1+(1-y^2)^2u^2(2-u^2)\alpha^2x^2)^2} \\ \nonumber\\
    W(y,u)&&=\frac{3-3y^2+y^4+(1-y^2)^2(1+u^2(2-u^2))\alpha^2-(1-y^2)^3u^2(2-u^2)\alpha^4}{1+(1-y^2)^2\alpha^2(2+\alpha^2-y^2(1+\alpha^2-u^2(2-u^2)(1-(1-y^2)\alpha^2)))}.
\end{eqnarray}
\end{widetext}
The benefit of inserting these functions into the \emph{Ansatz} is that the boundary conditions at the fictitious boundaries, at which we require regularity of the metric, become rather simple. At the axis of symmetry where the $S^2$ shrinks to zero size, $x=0$, we require that
\begin{eqnarray}
    \partial_x F_i(0,y,u)\, &&=  0 \quad \text{for } 1\leq i \leq 6 \nonumber \\
    F_i(0,y,u)\, &&=  0 \quad \text{for } 7\leq i \leq 9.
\end{eqnarray}

\noindent At the bifurcating Killing horizon, $y=0$, we have that
\begin{eqnarray}
    \partial_y F_i(x,0,u)\, &&=  0 \quad \text{for } 1\leq i \leq 6 \text{ or } i=9 \nonumber \\
    F_i(x,0,u)\, &&=  0 \quad \text{for } 7\leq i \leq 8.
\end{eqnarray}

\noindent The boundary conditions on the equatorial plane, $u=0$, are given by
\begin{eqnarray}
    \partial_u F_i(x,y,0)\, &&=  0 \quad \text{for } 1\leq i \leq 6 \text{ or } i=8 \nonumber \\
    F_i(x,y,0)\, &&=  0 \quad \text{for } i \in \{7,9\}.
\end{eqnarray}

\noindent Finally, at the north pole, $u=1$, we require that
\begin{eqnarray}
    \partial_u F_i(x,y,1)\, &&=  0 \quad \text{for } i \in \{1,2,4,5,6,8\} \nonumber \\
    F_i(x,y,1)\, &&=  0 \quad \text{for } i \in \{7,9\}\nonumber\\
    F_3(x,y,1)\,&&=F_4(x,y,1).
\end{eqnarray}

\bibliography{papers}

\end{document}